\documentstyle[epsf,twoside,fleqn,espcrc2]{article}
\title{Some aspects of the quark-antiquark Wilson loop formalism in the NRQCD framework}
\author{Nora Brambilla
\address{Institut f\"ur Theoretische Physik, Univ. Wien, Boltzmanngasse 5, A-1090 Vienna, Austria}
\thanks{Marie Curie fellow, TMR contract n. ERBFMBICT961714}        
and Antonio Vairo
\address{Institut f\"ur Hochenergiephysik, \"Oster. Akad.  der Wiss.,  
         Nikolsdorfergasse 18, A-1050 Vienna, Austria}}

\begin{document}

\begin{abstract}
Starting from the NRQCD Lagrangian the heavy quark-antiquark potential is written in terms 
of field strength insertions on a static Wilson loop. The relevant matching coefficients 
are given at the present status of knowledge. The short-range, perturbatively dominated, 
behaviour of the spin-dependent terms is discussed.
\end{abstract}

\maketitle

\section{NRQCD AND THE WILSON LOOP FORMALISM}

Heavy quark bound states provide an extremely difficult but at the same time 
appealing system to test QCD. The difficulties are obvious. 
One is the mixing of different energy scales. This is a typical feature of any 
bound state problem in quantum field theory and makes tricky even 
a purely perturbative solution of it. An other conceptual difficulty is connected with the 
nonperturbative nature of low-energy QCD. This suggests that nonpertubative contributions 
have to be taken into account in almost all QCD bound states.
The reason why heavy mesons are appealing is that the existence of an 
expansion parameter (the inverse of the mass $m$ in the Lagrangian and the 
velocity $v$ of the quark as a dynamical defined power counting parameter) 
makes possible to handle the first difficulty and to keep under 
control the second one. The tool is provided by NRQCD \cite{lepage}. 
This is an effective theory equivalent to QCD and obtained from 
QCD by integrating out the hard energy scale $m$. The Lagrangian comes from the 
original QCD Lagrangian via a Foldy--Wouthuysen transformation. The 
ultraviolet regime of QCD (at energy scale $m$) is perturbatively 
encoded order by order in the coupling constant $\alpha_{\rm s}$ 
in the matching coefficients which appear in front of the new operators 
of the effective theory. This ensures the equivalence between the effective  
theory and the original one at a given order in $1/m$ and $\alpha_{\rm s}$. 
At order $1/m^2$ the NRQCD Lagrangian describing a bound state between 
a quark of mass $m_1$ and an antiquark of mass $m_2$ is \cite{manohar,pineda}
\begin{eqnarray}
& & \!\!\!\!\!\!\!\!\!\!\!\! L = Q_1^\dagger\!\left(\!iD_0 + c^{(1)}_2 {{\bf D}^2\over 2 m_1} + 
c^{(1)}_4 {{\bf D}^4\over 8 m_1^3} + c^{(1)}_F g { {\bf \sigma}\cdot {\bf B} \over 2 m_1} \right.
\nonumber\\
& & \!\!\!\!\!\!\!\!\!\!\!\! + c^{(1)}_D g { {\bf D}\!\cdot\!{\bf E} - {\bf E}\!\cdot\!{\bf D} \over 8 m_1^2} 
\left.  \! + i c^{(1)}_S g {{\bf \sigma} \!\!\cdot \!\!({\bf D}\!\times\!{\bf E} - {\bf E}\!\times\!{\bf D})
\over 8 m_1^2} \!\right)\!Q_1
\nonumber\\
& & \!\!\!\!\!\!\!\!\!\!\!\!
+ \hbox{\, antiquark terms}\, (1 \leftrightarrow 2)
\nonumber\\
& & \!\!\!\!\!\!\!\!\!\!\!\!+ {d_1\over m_1 m_2} Q_1^\dagger Q_2 Q_2^\dagger Q_1 
+ {d_2\over m_1 m_2} Q_1^\dagger {\bf \sigma} Q_2 Q_2^\dagger {\bf \sigma} Q_1
\nonumber\\
& & \!\!\!\!\!\!\!\!\!\!\!\!+ {d_3\over m_1 m_2} Q_1^\dagger T^a Q_2 Q_2^\dagger T^a Q_1 
\nonumber\\
& & \!\!\!\!\!\!\!\!\!\!\!\!
+ {d_4\over m_1 m_2} Q_1^\dagger T^a {\bf \sigma} Q_2 
Q_2^\dagger T^a {\bf \sigma} Q_1.
\label{nrqcd}
\end{eqnarray}
This is the relevant Lagrangian in order to calculate the bound state 
observables up to order $O(v^4)$. A discussion of the operators appearing in 
(\ref{nrqcd}) in terms of powers of the quark velocity can be found 
in \cite{lepage,vairo}. The coefficients $c^{(j)}_2$, $c^{(j)}_4$, ... 
are evaluated at a matching scale $\mu$ for a particle of mass $m_j$. 

Nonperturbative contributions to the heavy meson observables can be evaluated 
directly from the Lagrangian (\ref{nrqcd}) via lattice simulations \cite{lepage}. 
Typically, since the hard degrees of freedom have been integrated out explicitly, 
the needed lattice cut-off $\mu$ is expected to be larger (smaller in terms of energy) 
than the usual one with a clear reduction in the computation time.  
Despite the advantages, there are also some drawbacks in this method. 
In particular in this way we do not learn very much on our "analytic" 
knowledge on the QCD vacuum structure. Moreover computations on coarse lattices  
are not always under control. Therefore it is worthwhile to use the  
Lagrangian of NRQCD as a starting point and to work out the quark-antiquark 
interaction in the so-called Wilson loop formalism \cite{brown}. 
The advantage in doing so is that all the nonperturbative dynamics 
will be contained in gauge field averages of field strength insertions 
on a static Wilson loop. These can be very easily evaluated by means 
of some QCD vacuum model \cite{bv}, or via traditional lattice simulations 
\cite{bali} providing in this way a powerful method in order to discriminate 
between different models. The derivation of the quark-antiquark potential  
in the Wilson loop formalism from the NRQCD Lagrangian was first suggested 
in this context in \cite{chen} and is discussed with details in \cite{vairo}. 
Here we present only some results. The heavy quark-antiquark potential  
(assumed that it exists) is given by
\begin{eqnarray}
& & \!\!\!\!\!\!\!\!\!\!\!\!
V(r) = \lim_{T \to \infty} { i \log W \over T} 
\nonumber\\
& & \!\!\!\!\!\!\!\!\!\!\!\!
+ \left( {{\bf S}^{(1)}\cdot{\bf L}^{(1)}\over m_1^2} +  
{{\bf S}^{(2)}\cdot{\bf L}^{(2)}\over m_2^2} \right)\!
{2 c^+_F V_1^\prime(r) + c^+_S V_0^\prime(r) \over 2r} 
\nonumber\\ 
& & \!\!\!\!\!\!\!\!\!\!\!\!
+ { {\bf S}^{(1)}\cdot{\bf L}^{(2)}  + 
{\bf S}^{(2)}\cdot{\bf L}^{(1)}  \over m_1 m_2} {c^+_F V_2^\prime(r) \over r}
\nonumber\\
& & \!\!\!\!\!\!\!\!\!\!\!\!
+ \left( {{\bf S}^{(1)}\cdot{\bf L}^{(1)}\over m_1^2} - 
{{\bf S}^{(2)}\cdot{\bf L}^{(2)}\over m_2^2} \right) \!
{2 c^-_F V_1^\prime(r) + c^-_S V_0^\prime(r) \over 2r} 
\nonumber\\
& & \!\!\!\!\!\!\!\!\!\!\!\!
+ { {\bf S}^{(1)}\cdot{\bf L}^{(2)}  - 
{\bf S}^{(2)}\cdot{\bf L}^{(1)}  \over m_1 m_2} {c^-_F V_2^\prime(r) \over r}
\nonumber\\
& & \!\!\!\!\!\!\!\!\!\!\!\!
+{1\over 8}\left( {c_D^{(1)} \over m_1^2} 
+ {c_D^{(2)} \over m_2^2} \right) (\Delta V_0(r) + \Delta V_a^E(r)) 
\nonumber\\
& & \!\!\!\!\!\!\!\!\!\!\!\!
+{1\over 8}\left( {c_F^{(1)} \over m_1^2} 
+ {c_F^{(2)} \over m_2^2} \right) \Delta V_a^B(r)
\nonumber\\
& & \!\!\!\!\!\!\!\!\!\!\!\!
+{c_F^{(1)}c_F^{(2)}\over m_1 m_2} \left( 
{{\bf S}^{(1)}\cdot{\bf r} {\bf S}^{(2)}\cdot{\bf r} \over r^2} - 
{{\bf S}^{(1)}\cdot {\bf S}^{(2)} \over 3} \right) \! V_3(r) 
\nonumber\\
& & \!\!\!\!\!\!\!\!\!\!\!\!
+ {{\bf S}^{(1)}\cdot {\bf S}^{(2)} \over 3 m_1 m_2}
\left( c_F^{(1)} c_F^{(2)} V_4(r) -48 \pi \alpha_{\rm s} C_F \, d \, \delta^3(r)\right)
\label{pot}
\end{eqnarray}
The ``potentials" $V_1$, $V_2$, ... are scale dependent gauge field averages of 
electric and magnetic field strength insertions on the static Wilson loop  
and are explicitly given in \cite{vairo,bali}. $W$ is the gauge average of the 
non-static Wilson loop. The expansion of it around the static Wilson loop gives 
the static potential $V_0$ plus velocity (non-spin) dependent terms. 
${\bf S}^{(j)}$ and ${\bf L}^{(j)}$ are the spin and orbital angular 
momentum operators of the particle $j$. The matching coefficients  
are defined as $2 c^{\pm}_{F,S} \equiv c^{(1)}_{F,S} \pm c^{(2)}_{F,S}$ and 
$d$ is the relevant contribution to the mixing coming from the 
four quark operators in Eq. (\ref{nrqcd}) and will be given in the 
next section. Apart from the matching Eq. (\ref{pot}) is equivalent 
to the potential derived in \cite{brown}. In the next section we 
will give explicitly the matching coefficients and discuss briefly the relevance 
of the matching in order to have a short range consistent potential.

\section{MATCHING COEFFICIENTS}

Since for reparameterization invariance $c_S = 2 c_F -1$ \cite{manohar},  
all the spin dependent potentials given in Eq. (\ref{pot}) turn out to depend only 
on $c_F$ (if the mass of the particle is irrelevant we will omit to indicate it).  
This coefficient is known up to two loop in the anomalous dimension \cite{amoros}:
\begin{eqnarray}
c_F &=& \left( {\alpha_{\rm s}(m)\over \alpha_{\rm s}(\mu) } \right)^{\gamma_0/2\beta_0}
\left[ 1 + {\alpha_{\rm s}(m)\over 4\pi} c_1 \right.
\nonumber\\
&+& \left. {\alpha_{\rm s}(m) - \alpha_{\rm s}(\mu) \over 4\pi} 
{\gamma_1\beta_0 - \gamma_0\beta_1 \over 2 \beta_0^2} \right]
\nonumber
\end{eqnarray}
where $\beta_j$ are the usual $\beta$-function coefficients, $\gamma_0 = 2 C_A$, 
$\gamma_1 = 68 C_A^2/9 - 26 C_A N_f/9$, $c_1 = 2 (C_A + C_F)$, $N_f$ is the number of flavors, 
$C_F$ is the Casimir of the fundamental representation and  $C_A$ is the Casimir 
of the adjoint representation. At the lattice scale used in \cite{bali} the numerical 
values of this coefficient at the bottom and charm mass are $c_F(m_b) \simeq 1.06 \times (1 + 0.15) = 1.22$ 
and $c_F(m_c) \simeq 1.27 \times (1 + 0.25) = 1.59$ respectively. The corrections due to the 
one loop matching are relevant (15 \% in the bottom case and 25 \% in the charm case) and 
therefore of the same order of the next power in the velocity in the Lagrangian (\ref{nrqcd}) 
(usually accepted values are $\langle v^2_b \rangle \sim 0.07$ and $\langle v^2_c \rangle \sim 0.24$). 

An evaluation of the coefficient $c_D$ associated with the Darwin term in the NRQCD Lagrangian 
is given in \cite{balzereit}:
\begin{eqnarray}
c_D &=& \left({7\over4} - 8 {C_F\over C_A} \right) \left( {\alpha_{\rm s}(m)\over \alpha_{\rm s}(\mu) } 
\right)^{2 C_A / 3\beta_0} 
\nonumber\\
&-& {5 \over 4} \left( {\alpha_{\rm s}(m)\over \alpha_{\rm s}(\mu) } 
\right)^{2 C_A / \beta_0} + {1\over 2} + 8 {C_F\over C_A}.
\nonumber
\end{eqnarray}
This corrects a previous wrong evaluation given in \cite{chen}. 
At the lattice scale used in \cite{bali} the numerical values of this coefficient 
at the bottom and charm mass are $c_D(m_b) \simeq 0.76$ and $c_D(m_c) \simeq  -0.08$ respectively.
As pointed out in \cite{bali}, since the potential $\Delta V_a^E$ manifests a $1/r$ behaviour, 
this term gives a flavor-dependent contribution to the central potential. 
However this contribution is suppressed in the bottom case by the bottom mass (see Eq. (\ref{pot})) 
and in the charm case by the smallness of the corresponding matching coefficient. 

Finally, the contributions coming from the four-fermion operators are usually suppressed 
either in $\alpha_{\rm s}$ or in powers of the quark velocity $v$ \cite{lepage,pineda,vairo}. 
Nevertheless under RG transformation the contribution to the spin-spin potential 
coming from the chromomagnetic operator in the NRQCD Lagrangian mixes with some of the  
local four quark operators. In order to take into account this mixing the delta contribution 
to the spin-spin potential has been added in Eq. (\ref{pot}) though it would be  
suppressed in $\alpha_{\rm s}$. The coefficient $d$ has been evaluated in \cite{chen}:
$$
d = {1\over 8} \!\left( {\alpha_{\rm s}(m_1)\over \alpha_{\rm s}(m_2) } 
\right)^{C_A / \beta_0} \!\! \left[1 - \left( {\alpha_{\rm s}(m_2)\over \alpha_{\rm s}(\mu) } 
\right)^{2 C_A / \beta_0} \right].
$$

\begin{figure}[htb]
\makebox[3.5truecm]{\phantom b}
\epsfxsize=7.5truecm
\epsffile{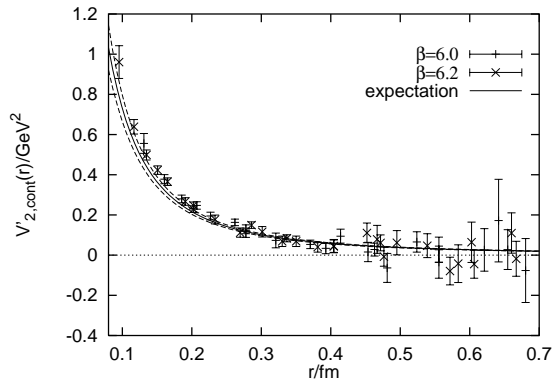}
\vskip -1 truecm 
\caption{The spin dependent potential $V_2^\prime$ as given by the lattice 
         measurement of \cite{bali}.}
\label{figv2lat}
\end{figure}

\begin{figure}[htb]
\makebox[3.5truecm]{\phantom b}
\epsfxsize=7.5truecm
\epsffile{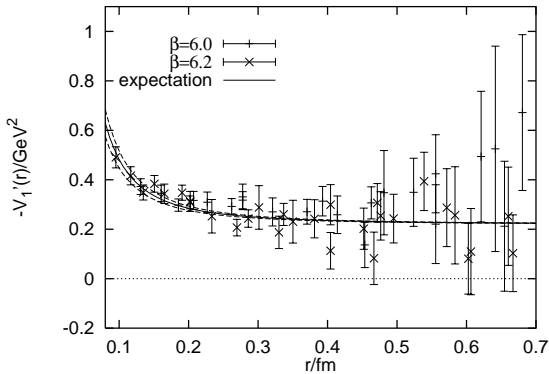}
\vskip -1 truecm
\caption{The spin dependent potential $-V_1^\prime$ as given by the lattice 
         measurement of \cite{bali}.}
\label{figv1lat}
\end{figure}

As noticed in \cite{chen} the presence of the matching coefficients in the expression  
for the potential makes possible the agreement in the short range region between the potential 
derived here with the traditional QCD one loop perturbative calculation, e.g. in \cite{ng}.  
Let us focus on the spin-orbit terms $V_1$ and $V_2$. Comparing properly with \cite{ng} 
we get for $V_2$ the  perturbative contribution 
\begin{eqnarray}
V_{2,pert}^\prime(r) &=& {C_F\alpha_{\rm s}(\mu)\over r^2} 
\left\{ 1 + {\alpha_{\rm s} \over \pi} \left[ - {\beta_0 \over 12} - {2C_A\over 3}
\right.\right.
\nonumber\\
&+& \left.\left.  {\beta_0 - C_A \over 2} (\log (\mu r) + \gamma_E -1) \right]\right\}, 
\nonumber
\end{eqnarray}
where $\gamma_E$ is the Euler constant. This expression agrees very well with the lattice 
measurement of the same quantity shown in Fig. \ref{figv2lat}. 
In the same way we get for $V_1$ the  perturbative contribution 
$$
V_{1,pert}^\prime(r) = - {\alpha_{\rm s}^2 \over \pi} {1\over r^2} {C_A C_F\over 2}
(\log (\mu r) + \gamma_E). 
$$
It is extremely interesting to compare the above expression with the short-range behaviour 
of the $V_1$ potential as given by the lattice measurement shown in Fig. \ref{figv1lat}. 
Apart an overall shift proportional to the string tension and therefore of nonperturbative 
origin the agreement is very good. This is quite significant since the perturbative part of 
$V_1$ is entirely  due to loop corrections. As a consequence $V_1$ is more sensitive than $V_2$ 
to the matching scale $\mu$. Notice that at very short distances the function  $-V_1^\prime$,  
just because the $\log (\mu r)$ term, is expected to become negative, but up to now no lattice 
data are available in this region.
As a last comment we notice that due to the so-called Gromes relation 
$V_2^\prime - V_1^\prime = V_0^\prime$ \cite{gromes} a $V_5$ potential (in the notation of 
\cite{ng}) emerges also in Eq. (\ref{pot}) by collecting the contributions 
coming from the fourth and fifth line. The perturbative expression we get agrees with 
that one given in \cite{ng}:
$$
V_{5,pert}(r) = {c_F^- V_2^\prime(r) \over r} = {\alpha_{\rm s}^2 \over \pi} 
{1 \over r^3} {C_A C_F \over 4} \log {m_2 \over m_1} .
$$

\section{CONCLUSIONS}

In the framework of NRQCD and at the present status of the matching we have 
given the expression for the heavy quark potential in terms of field strength 
insertions on a static Wilson loop. This has 
the advantage that traditional lattice calculations can be used in order 
to evaluate nonperturbative contributions. Moreover in this way a 
comparison between different QCD vacuum models can be performed directly 
in terms of Wilson loop expectation values. This approach  has been developed 
with some extent in \cite{vairo}. Here we have emphasized the role played by the 
matching coefficients in order to make consistent the short range behaviour 
of the potential that we obtain with the usual scattering matrix derived potential. 
We noticed that present lattice data are sensitive to one loop corrections and to 
the matching scale. 

As a conclusion, let us mention two open problems. In order to have a 10 \% accuracy 
in the quarkonium spin splitting it is necessary to add to the Lagrangian (\ref{nrqcd}) 
higher order operators \cite{lepage}. The  inclusion (if possible) of  such 
operators in an expression like Eq. (\ref{pot}) is still to do. Moreover in order to obtain 
Eq. (\ref{pot}) we have implicitly assumed the existence of a potential. 
Non-potential terms surely exist in perturbative QCD. How to treat it 
in a system affected by nonperturbative physics is still unclear. Interesting 
developments could come from a promising approach recently proposed in \cite{pnrqcd}.

{\bf Acknowledgments} We thank Antonio Pineda for valuable discussions.

\end{document}